\documentstyle[aps,preprint]{revtex}
\draft

\begin{document}
\title{ Ginzburg-Landau Description of Vortex Nucleation in Rotating Superfluid.}
\author{Igor Aranson$^a$ and Victor Steinberg$^b$}
\address{$^a$Argonne National Laboratory, 9700 South Cass Avenue,\\
Argonne, Illinois 60439, USA\\
$^b$ Department of Physics of Complex Systems, Weizmann Institute of\\
Science, Rehovot, 76100, Israel}
\date{\today}
\maketitle

\begin{abstract}
Nucleation of vortices in rotating superfluid 
by spin-up and rapid thermal quench is discussed in the framework of 
the time-dependent Ginzburg--Landau equation (TDGLE). 
An analysis of the instability 
in inhomogeneous
rotationally-invariant system results in the expression for the critical rotational 
velocity. 
A stability analysis of multicharged
vortices is presented. 
It is shown that they 
are very long-living objects with lifetime inversely
proportional to the dissipation rate.
It was found by 
numerical and analytical solution
of the  
TDGLE that vortex nucleation 
by rapid thermal quench in the presence of superflow
is dominated by a transverse instability of the moving
normal-superfluid interface.

\end{abstract}
\newpage
\section {Introduction.}
There are two major ways to generate vorticity in rotating superfluid: 
to spin-up a bucket with 
superfluid helium from initially steady state
 without any vortices to a state with a rotation 
velocity $\Omega\geq \Omega_c$, where $\Omega_c$ 
is the critical rotation speed for a vortex
nucleation, and to quench it thermally from temperature, $T$, 
above the superfluid 
transition temperature, $T_{\lambda}$, to temperature below it. 
These two different scenaria
can be described by the same model, 
namely time-dependent Ginzburg--Landau 
equation (TDGLE)
for the superfluid order parameter corrected for 
presence of normal component. This equation for 
superfluid helium was first suggested by 
L. Pitaevskii\cite{ginz,pit}.

Further we review our results 
based on the stability analysis of the TDGLE  for the rotating
superfluid. 
The vortex nucleation in a rotating superfluid was 
a rather hot topic during 60th and 70th. 
Remnant vorticity did not permit to define correctly the 
critical rotational velocity for vortex nucleation 
at that time. We believe that next
generation of experiments will overcome this 
problem, e.g., in the way suggested about 25 years 
ago\cite{hulin}. Together with new experimental 
methods of vortex detection suggested 
recently\cite{lund,david}, 
it can provide experimental test of the theory.

Formation of vortices  under
a rapid quench is recognized as a fundamental problem of
contemporary physics \cite{volovik}.
Superfluid $^3$He offers a unique ``testing ground''  for
rapid phase transitions \cite{ekv}. 
Recent experiments where a rotating superfluid $^3$He was locally heated
well above the
critical temperature by absorption of neutrons \cite{explos}
revealed vortex formation under a rapid second--order phase transition.
We will discuss the dynamics of vortex nucleation 
under a rapid quench in the framework of the TDGLE.

\section{Spin-up and nucleation of vortices in superfluid helium}

Let us consider a cell containing a superfluid helium rather close to the
superfluid transition temperature $T_\lambda $. When the cell is rotated
with an angular velocity $\Omega $, the normal component is involved into a
solid body rotation with $\vec {V_n}=\vec \Omega \times \vec r$. 
The superfluid component 
cannot participate in the
uniform rotation up to the point where the potential flow condition is
satisfied, i.e. $\vec \nabla \times \vec {V_s}=0$.

A conventional approach to the spin-up problem is to describe it by a
two-fluid hydrodynamic model corrected by an equation for the vortex line
dynamics \cite{reisn,don}. The vortex lines 
interact with the normal component that leads to mutual friction\cite
{don}. It is evident that this hydrodynamic description does not catch the key
point of the spin-up problem, namely, the vortex nucleation which actually
causes the spin-up of the superfluid component, 
the primary superfluid
relaxation mechanism toward a steady rotation.

Another approach to the spin-up problem, 
which can describe both the dynamics and the nucleation of quantized 
vortices in a
superfluid helium, is to use TDGLE 
together with a two-fluid hydrodynamic model\cite{pit}.
Outside the vortex
core which is normal, one gets the superfluid velocity circulation around a single quantum vortex 
$\kappa =2\pi \hbar/m $ \cite{don}.
Corresponding set of equations, which 
describe the dynamics of the complex order parameter of the
superfluid condensate, $\Psi=|\Psi|\exp(i\chi)$, in a rigid steady rotation of the normal component, 
looks in the scaled variables as follows \cite{pit}( see for details 
\cite{aran}):
\begin{eqnarray}
\partial _t\Psi =-\frac i2\left( \Delta \Psi +\Psi -|\Psi |^2\Psi\right) 
+\frac \Lambda 2\left( (\nabla -i{\bf V_n})^2\Psi +\Psi -|\Psi |^2\Psi
\right)  \label{eqno1} 
\end{eqnarray}
where $V_n=\Omega r$,  ${\bf V_s}=\nabla\chi$, and $\Lambda$ is a
temperature-dependent  parameter.

Equation (\ref{eqno1}) 
is reminiscent of that of the Ginzburg--Landau equation
for superconductors in the London limit   \cite{super}. The role
of an external magnetic field is played by the angular velocity, and of the
corresponding vector-potential by the velocity of the normal component.
Then by analogy one expects that at $\Omega\leq\Omega_{c1}$, there exists a
motionless superfluid component with no vortices. At $\Omega > \Omega_{c1}$
vortices will be nucleated and penetrate into the fluid producing a vortex
lattice in the interior of a helium container. As follows from the
experiments on the superfluid $^4$ He, $\Omega_{c1}$
 is too low to be detected
\cite{don}. 
On the other hand, $\Omega_{c2}$ (which is analogous to $H_{c2}$
in superconductors and at which superfluidity will be completely destroyed
in the sample) is too high to be reached experimentally.

Equation (\ref{eqno1})
describes  the spin-up of the superfluid 
part in a rigidly rotating flow of the
normal component. 

This equation is asymptotically correct in the vicinity of
the $\lambda $ point. Moreover, one can speculate that the equation
qualitatively  describes some aspects of spin-up for $^4 He$ near zero temperature with 
$\Lambda \to 0$. 
Therefore, we will consider Eq. (\ref{eqno1})
for all temperatures with  $\Lambda \to 0$ for $T\to 0$ 
and  $\Lambda\sim (T_\lambda-T)^{-1/3}$ for $T\to T_\lambda $.

The critical rotational velocity, $\Omega_c$, 
for the onset of the vortex nucleation for $T \to T_\lambda$ can be found from 
linear stability analysis of a stationary solution.
Then for this solution $\chi=0$, and 
$F=|\Psi|$ is defined by the following equation: 
\begin{equation} 
\partial _r^2F+\frac{\partial _rF}r+F-F^3-\Omega ^2r^2F=0. 
\label{eq4}
\end{equation} 
Equation (\ref{eq4}) has to be complimented by the conditions
at the $r=0$ and the condition at the outer  wall $r=R$,
where $R$ is the radius of the container. 
As a boundary condition at the wall we
take a condition of finite suppression of the superfluid density by the
wall, i.e., 
$\partial _r\Psi +\gamma \Psi =0$  for $r=R$
where $\gamma $ characterizes the suppression of the order parameter. 
For $
\gamma \to 0$ we have no-flux boundary condition ($\partial _r\Psi =0$).
Solution of Eq. (\ref{eq4}) for arbitrary  $\Omega$ and $R$ is accessible only
numerically. Selected results are presented 
in Fig.1 of Ref\cite{aran}.

The stationary solution is stable for $\Omega<\Omega_c$ and looses its  
stability
above the critical angular velocity $\Omega_c$. 
Instability of the stationary solution leads to nucleation of
vortices and the corresponding spin-up of superfluid.
By substitution of a solution of the form $\Psi = F(r) + W(r,\theta,t)$,
where $W$ is a small generic perturbation, one obtains a 
linear equation for $W$. 
$\Omega_c$ is found from the existence condition of the 
first nontrivial eigenmode
satisfying the boundary conditions \cite{aran}. 
In the limit of $R\gg 1$ the solution is found by 
matching  of the bulk solution with the solution near the wall.
The analysis shows that the most unstable eigenmodes are localized in the
narrow layer of the width $r_b$ near the container wall. We obtained $%
r_b\sim \sqrt{R}\ll R$ for large $R$. The value of most unstable azimuthal
number $n$ and the critical frequency $\Omega_c $ for the container radius $R$
is given by the expressions: 
\begin{equation} 
n =Q(\gamma )R^{3/4},\;\;\Omega_c =\frac 1R\sqrt{\frac 13+
\frac{\Delta (\gamma )}{R^{1/2}}}.
\label{eq3}
\end{equation} 
The dimensionless parameters $Q(\gamma ),\Delta (\gamma )$ are the functions
of the suppression rate $\gamma $, are obtained by the matching of outer
and inner expansions, and are shown in Fig. 2 of Ref \cite{aran}.
One can make estimates of $\Omega_c$ based on Eq.(3). Indeed, in dimensional
variables one gets
$\Omega_c=(\hbar/ m R_d)[\epsilon^{2/3}/(\sqrt{3}\xi_0 )]=3.3\times 10^3\epsilon^{2/3}/R_d$ 1/sec,
where $\epsilon=(T_{\lambda}-T)/T_{\lambda}$, $\xi_0=2.74\times 10^{-8}$ cm is the correlation 
length far from $T_{\lambda}$, and the power 2/3 
is introduced 
to assure correct scaling of the superfluid density near the $\lambda$ point \cite{ginzb}. 
The conventional Feynman equation \cite{don} has different scaling with $R$ and no temperature dependence.
On the other hand, as we discussed in Ref.\cite{aran}, the temperature dependence of $\Omega_c$ found is 
the same as
in the theory of
thermal nucleation of quantized rings \cite{don}. However, the nucleation rate in our case was calculated 
up to prefactor (see Ref.\cite{aran}), while in the former theory it was obtained from heuristic arguments.
Moreover, we consider non-uniform distribution of $\rho_s$ due to $V_n$ that is impossible task for the 
equilibrium theory. The latter can become significant at small values of $\epsilon$ and $R$. Thus, at 
$\epsilon =10^{-6}$ and $R_d=0.1cm$ scaled $R$ is of the order $10^3$,
and the correction to the frequency from Eq.(3)
can be of the order of several percent.

We performed numerical simulations of Eq.(\ref{eqno1}),
which details are presented in Ref\cite{aran}. 
We observed nucleation and consequent tearing off of the vortices for $
\Omega \ge \Omega_c$ irrespective of $\Lambda$. However, the character of
the nucleation and asymptotic states depends  on $\Lambda$.
For  $\Lambda \to \infty $ and  slightly above 
$\Omega _c$ we observed nucleation of several vortices.
Nucleation occurs  at nonlinear stage of the instability when a set of single
zeros (four zeros for $R=65$) is torn off at the radius $R$. These zeros
are the seeds for the vortex cores. The vortices propagate into the interior
of the container and finally form a perfect vortex lattice, reminiscent of that of the
Abrikosov lattice \cite{super} (see Fig. 1). Further increase of $\Omega $
results in formation of additional vortices.

\section{Stability of multicharged vortices}

As we pointed out above, Eq.(\ref{eqno1}) can describe qualitatively 
the spin-up of superfluid helium also at $T\to 0$ with
$\Lambda \to 0$. Then the TDGLE becomes the nonlinear 
Sch\"{o}dinger equation (NLSE).
Numerical simulations show  the character of the vortex nucleation is 
drastically 
different for $\Lambda \to 0$. Typically  
whole clusters of vortices are torn off. These clusters 
can be considered as a perturbed multicharged
vortex. The multicharged vortex is unstable and breaks down into
single charged vortices. However, the lifetime 
happens to be proportional to $\Lambda$ and diverges for $\Lambda \to 0$.
The multicharged
vortices with the topological charge 
$\pm n$ are known to have higher energy and decay into $n$ 
single-charged, or elementary vortices. Since in the NLSE 
with small dissipation the decay
time of the multicharged vortices can be arbitrarily large, 
one can expect to detect them in experiments at 
very low $T$. Some indirect indications of 
the multicharged vortices can be found in several 
experiments on vortex nucleation \cite{don}.

We consider the 
perturbative solution $\Psi=[F(r)+\eta(x,y,t)]\exp[in\theta]$ of the NLSE with dissipation\cite{aran1}
\begin{equation} 
\partial _t\Psi =(\varepsilon +i)\left( \Delta \Psi +\Psi -|\Psi |^2
\Psi \right), \label{eq11}
\end{equation} 
Here $\eta$ is the complex function, and $\varepsilon \ll 1$ 
is the phenomenological parameter which describes
the bulk dissipation of superflow towards the condensate. 
We assume that the only channel for the bulk 
dissipation at $T \to 0$ is the absorption of 
acoustic excitations of superflow by the normal component which
presents, e.g., due to normal $^3$He atoms.
 This assumption is based on consideration that the excess energy 
of $n$-charged vortex can decay at $T\to 0$ 
only by acoustic radiation of the bounding energy of $n$ single
charges. In the presence of the energy conservation and other 
integrals of motion in the NLSE this 
transformation of the bounding energy into 
acoustic field is a very slow process, i.e.
multiple vortex can be  long-lived.
Numerical analysis of the stability problem reveals that  the
lifetime of the multicharged vortex 
diverges as $\tau\sim (\varepsilon\lambda_1)^{-1}$ and is formally infinite for 
NLSE. 
However, one has slower (non-exponential) instability mode at $\epsilon=0$. 
In particularly, 
generic perturbations grow linearly in time. 
In this sense the multicharged
vortex is metastable and may exist for a very long time. 
The instability has a nonlinear nature and originates 
from the interaction between localized eigenmodes and
the continuous spectrum of the vortex radiation. 
The crucial point here is that the simultaneous existence 
of the localized eigenmodes with $n\geq 2$ and extended 
excitations does not contradict to the energy 
conservation of the Hamiltonian NLSE, 
since they contribute to the energy with opposite signs. This
process is similar to the growth of waves with negative energy \cite{ostr}. 
In Fig.2 one clearly sees that
a rotating double-charged and triple-charged vortices 
radiate away the acoustic waves. Thus, the decaying 
multicharged vortices are effective source of the acoustic radiation.

\section{Nucleation of vortices by rapid thermal quench}
Nucleation  of vortices  by neutron irradiation in 
$^3$He-B in the presence of 
rotation  was studied experimentally in Ref. \cite{explos}. 
Ignoring non-relevant complexity of the $^3$He-B specific
multicomponent order parameter, we will use the TDGLE for 
a  scalar order parameter $\psi$ \cite{akv}: 
\begin{equation}
\partial_t \psi = \Delta \psi + (1 - f({\bf r },t) ) \psi -|\psi| ^2 \psi
+\zeta({\bf r},t).
\label{gle1}
\end{equation}
Close to $T_\lambda$
the local temperature is controlled by normal-state heat diffusion and
evolves as
$f({\bf r },t) =  E_0  \exp(-r^2 /\sigma t)t^{-3/2}$, 
where $\sigma$ is the normalized diffusion coefficient.
$E_0 \gg 1 $ determines the initial temperature of the hot
bubble $T^*$ due to the nuclear reaction between the neutron and $^3$He atom 
and is
proportional to the deposited energy ${\cal E}_0$. 
The Langevin force
$\zeta$  with the correlator
$\langle \zeta \zeta^\prime \rangle=
2 T_f \delta({\bf r -r^\prime}) \delta(t-t^\prime)$
describes  thermal fluctuations
with a  strength $T_f$. 

Numerical simulations of Eq. (\ref{gle1}) in 2D and 3D shows 
that without fluctuations ($T_f=0$) 
 the  vortex rings nucleate upon the passage of the
thermal front, Fig. 3a,b. Not all of the rings survive: the
small ones collapse and only the big ones grow. Although the vortex lines
are centered around the  point of the quench, they exhibit a
certain degree of entanglement. After a long transient period,
most of the vortex rings reconnect and form the
almost axisymmetric configuration.
We find that the fluctuations have a  strong effect at
early stages:
the vortices nucleate not only at the
normal-superfluid interface, but also in the bulk of the supercooled
region (Fig. 3c), according to the Kibble-Zurek ``cosmological'' mechanism \cite{volovik}.
However,  later on, small vortex rings in the interior collapse, and only
larger rings
(primary vortices) survive and expand (Fig. 3d).

We conclude that the primary source of vortices 
in our numerical simulations is the instability of normal-superfluid 
interface in the presence of flow. The analysis 
results in the following expression for the number of survived vortices 
\begin{equation}
N   \sim  \sqrt\sigma E_0^{1/3}\sqrt{ \left(v_s/v_c\right)^2
-\beta^2\log(T_f^{-1})/ E_0^{2/3} }
\label{estimate}
\end{equation}
where $\beta=const$, while $v_s$ and $v_c$ are the imposed and critical GL
superflow velocity, respectively. This estimate is in agreement with
simulations, see Ref. \cite{akv}. 
Eq. (\ref{estimate}) exhibits a slow logarithmic dependence of the
number of vortices 
on the level of fluctuations. 
For
 the experimental
values of the parameters
our analysis results in about 10
surviving vortices per heating event. It is consistent with Ref. \cite{explos}
where as many as 6-20 vortices per neutron were detected.

This research is supported by US DOE, grant W-31-109-ENG-38 (I.A.), and by the Minerva 
Center for Nonlinear Physics of Complex Systems (V.S.).

\begin{figure}
\caption{The sequence of grey-coded images of $|\Psi|$  demonstrating nucleation of vortices and
creation of the vortex lattice. The dark shade corresponds to zero of $|\Psi|$; the white one
to its
maximum value. Vortices are seen as black dots. The parameters are: 
$\Omega =0.01, R=65, \Lambda \gg 1 $. The initial condition is $\Psi = 1$
plus small amplitude broad-band noise. a) $t=100 \Lambda $, b) $t=200 \Lambda $, c) $t=1100 \Lambda $. 
  }
\label{figa}
\end{figure}

\begin{figure}
\caption{Number of vortices as a function of $\Omega$. a)  $R=65 $; 
b) $R=140$.
}
\label{figb}
\end{figure}
\begin{figure}
\caption{Gray-coded snapshots of $|\Psi(x,y)|^2$(zero is shown in black,
$|\Psi|=1$ in white) for double-charged vortex(a)-(c); and triple-charged (d)-(f) at the moments
of time :
(a) t=1700; (b) t=2000; (c) t=2500;(d) t=1700; (e) t=2000; and (f) t=2500. The parameters of the 
simulations are: The domain size $100\times100$ units, number of FFT harmonics $128\times 128$;
$\varepsilon=0.001$, boundary conditions no-flux; initial conditions were slightly perturbd 
double-charged 
vortex. Single vortices are presented by black spots, the acoustic field is seen in gray shade.}

\label{figc}
\end{figure}

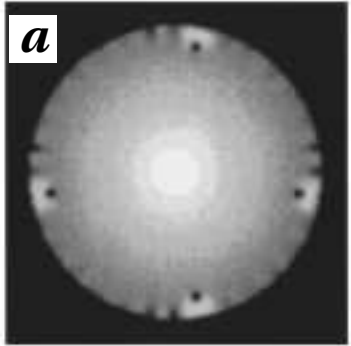
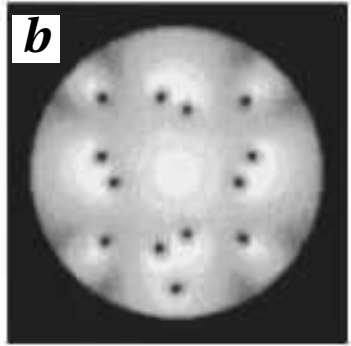
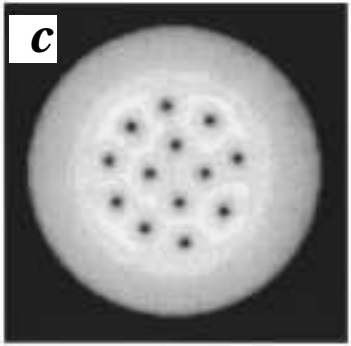
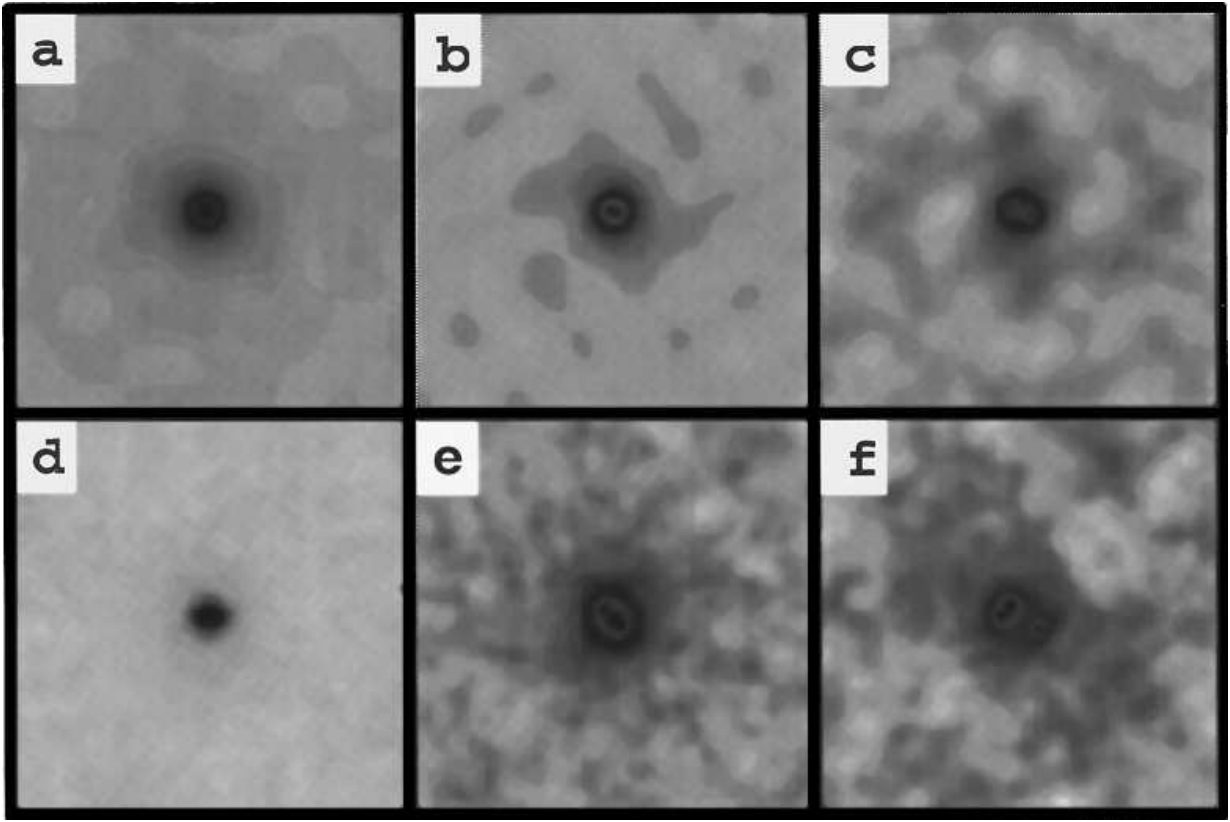
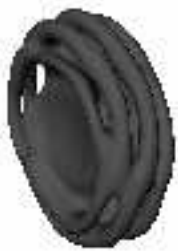
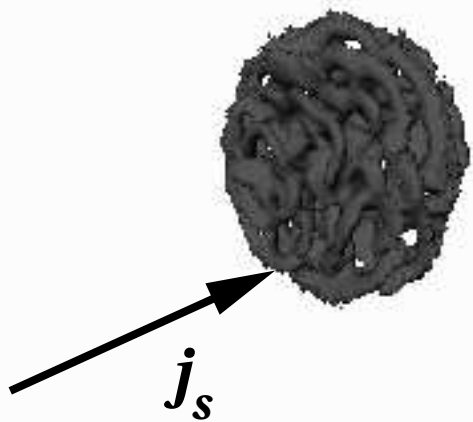
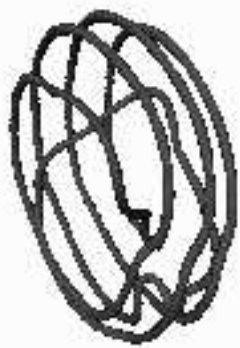
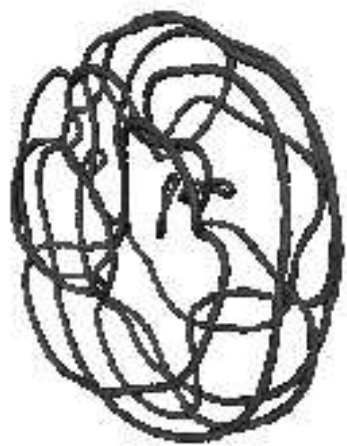
\end{document}